# Growth of *Phaseolus vulgaris* in Response to Seed Priming by Plasma-Activated Water in Laboratory Screening and Outdoor Pot Trial

Mustafa Ghulam[1], Ramin Mehrabifard[2], Adriana Mišúthová[2], Zuzana Lukačová[3], Pratik Doshi[1], Zdenko Machala[2], Božena Šerá[1]*

[1] Department of Environmental Ecology and Landscape Management, Comenius University Bratislava, Ilkovičova 6, 842 15 Bratislava, Slovakia

[2] Division of Environmental Physics, Faculty of Mathematics, Physics and Informatics, Comenius University Bratislava, Mlynská dolina, 842 48 Bratislava, Slovakia

[3] Department of Plant Physiology, Comenius University Bratislava, Ilkovičova 6, 842 15 Bratislava, Slovakia

Correspondence: Božena Šerá (bozena.sera@uniba.sk)

## Abstract

This study explores plasma-activated water (PAW) effects on Common bean growth in laboratory and pot trials. Three treatments were assessed: PAW priming, spraying, and their combination. Laboratory trials showed no germination improvement. However, pot trials revealed notable increases in seedling length, biomass, and antioxidant enzyme activity. Enzymes SOD, G-POX, CAT, APX, and GR showed significantly higher activity in PAW-treated plants. These effects were linked to reactive oxygen and nitrogen species in PAW. Findings suggest PAW enhances bean growth and physiology, supporting field farming applications.

**Keywords:** Plasma Activated Water (PAW), Seed Priming Phaseolus vulgaris (Common Bean) Cold Atmospheric Plasma (CAP), Reactive Oxygen and Nitrogen Species (RONS), Plant Growth and Antioxidant Response, Sustainable Agriculture / Biostimulants.

# 1. Introduction

Common bean (*Phaseolus vulgaris* L.) is of paramount importance due to food production as it is the most significant leguminous grain consumed by people (50% of the total leguminous grain consumption) [1]. It is one of the three most important legumes worldwide together with soybeans (*Glycine max* L.) and peanuts (*Arachis hypogaea* L.) [2]. Beans are the richest source of protein, fibre, minerals, slowly digestible starch, and several other minerals and nutrients, thus making beans a superfood [3]. Beans are grown in several eco-agricultural regions and are found in multiple forms, such as whole seeds, mixed products, canned items, or as a gluten-free substitute for wheat flour [4].

Common beans contribute significantly to food supplies and production and have important social and economic implications. There are several factors that can reduce bean yields. More than 45 diseases can badly impact and hinder bean growth during their life cycle [5]. The main diseases are anthracnose, angular leaf spot, and common mosaic, which effect the common beans plants and reduce their yield and quality [6]. Furthermore, beans can be affected by various biotic and abiotic environmental stresses, including drought, flooding, salinity, extreme temperatures, heavy metal exposure, and nutrient deficiencies [7].

In the past few decades, chemical fertilizers have been widely applied to enhance crop growth and yields. However, it has become evident that the use of chemical fertilizers has had very negative impacts on the environment, including various health hazards for animals, plants, and the overall ecosystem [8]. It is estimated that the world's population will reach the magnitude of 10 billion people by 2050 [9]. The need of the hour is to find solutions that are sustainable, environmentally friendly, and improve crop yields to satisfy the human needs.

Seed priming, which involves the controlled hydration of seeds to boost germination, has been utilized since ancient times. Historical accounts from Theophrastus and Roman authors refer to the practice of soaking seeds in water or milk to accelerate sprouting [10]. Contemporary research has enhanced this ancient method, demonstrating that priming triggers early metabolic and hormonal processes that enhance both germination and seedling vitality [11]. Recent advancements in this technology have shown its potential, which offers resistance against various biotic and abiotic stresses, as well as enhancement of the crop yield [12, 13] . Significant progress has been made in the field of seed priming, with researchers exploring modern techniques and approaches to optimize its full efficiency. It has gained increasing attention as an eco-friendly strategy that has been consistently shown to significantly improve germination percentage, seedling vitality, and overall crop condition and product yield [14].

Cold atmospheric plasma (CAP) has been used across several scientific disciplines owing to its exceptional characteristics. CAP produces active species (positive/negative ions, radicals, electrons), UV radiation, and transient electric fields [15–17]. Active species dissolve in water when CAP comes into contact with it and generate so-called plasma-activated water (PAW) [18].

PAW is being progressively identified as an emerging and environmentally sustainable technology in agriculture, showing considerable promise for enhancing seed germination, plant growth, and the biocontrol of pests and diseases. PAW is produced by exposing water to CAP, typically generated through various types of electrical discharges. The plasma–liquid interaction initiates complex chemical reactions both within the gas, typically air, the bulk liquid, typically water, and within the gas–liquid interface, thereby modifying the physicochemical properties of the liquid. The non-equilibrium plasma reactions result in the formation of various short- and long-lived reactive oxygen and nitrogen species (RONS), which are transported from the plasma into the liquid. Typical long-lived RONS in PAW are hydrogen peroxide ($H_2O_2$), ozone ($O_3$), nitrite ($NO_2^-$), and nitrate ($NO_3^-$) [19].

Reactive oxygen species (ROS) in seeds and plants mediate numerous intracellular signalling and metabolic pathways, while reactive nitrogen species (RNS) contribute mostly to nitrogen assimilation and availability. Nitrite and nitrate represent key bioavailable forms of nitrogen, an essential macronutrient for plant growth and metabolism. However, RNS with $H_2O_2$ function as signalling molecules activating proteins and genes involved in plant developmental processes and various stress reactions [20, 21]. The most important in this treatment is the optimal concentration of RONS: a delicate balance, maintaining their concentrations accurate to be neutralized by plant antioxidants, yet providing the stimulating effect.

The application of PAW – either through seed imbibition, foliar application, or irrigation – has been shown to promote and accelerate seed germination via hormonal modulation, enhance seedling vitality, increase photosynthetic pigment concentrations, and regulate antioxidant enzymes [14, 22–25]. Nevertheless, excessive accumulation of $H_2O_2$ and $NO_2^-$ may lead to oxidative stress, mimicking plant responses under various abiotic stress conditions [26]. Disruption of RONS homeostasis can trigger deleterious effects, including chlorophyll degradation, reduced photosynthetic efficiency, and impaired growth.

In response, plants activate a suite of defence mechanisms to counteract RONS-induced damage. These include the upregulation of non-enzymatic antioxidants (e.g., phenolic compounds, carotenoids), enzymatic antioxidants (e.g., peroxidases, catalase, glutathione reductase or superoxide dismutase), and structural defences, such as enhanced lignification [24, 26].

Seed priming with PAW initiates metabolic process in the surface of seeds and has direct effect in seedling germination and growth [27]. Several studies investigated the effect of seed priming with PAW [28]. Seeds of four distinct species: buckwheat (*Fagopyrum esculentum* Moench.), barley (*Hordeum vulgare* L.), black mustard (*Brassica nigra* (L.) W.D.J.Koch), and brown mustard (*Brassica juncea* (L.) Czern.) were primed with PAW, and showcased significant increase in germination parameters, seedling lengths and seedling vitality compared to control [29]. A study investigated peanut seeds, when primed with PAW, they showed significant improvement in early germination and improved biomass-related morphological traits in peanut sprout as compared to control [30].

In hydroponics settings, PAW induced priming significantly improved the germination rate (98.3% vs. ~80% in the control), germination uniformity (~5 hours vs. ~15 hours in the control), and reduced the mean germination time (~1.87 days vs. ~2.5 days in the control) in green oak lettuce [31]. Many studies show that, depending on the characteristics and types of plasma species generated during the PAW production, PAW priming is effective in improving drought tolerance, has antifungal and antibacterial properties, boosts yields, and enhances self-resistance mechanisms in plants [32]. Rasid et al. 2021 [32] investigated how foliar spray combined with direct seed application of on rice seeds (*Oryza sativa* L.) enhanced plant growth parameters. Combined application resulted in improved defence mechanism through enhanced enzymatic activities, while concentrations of total soluble protein and sugar were also increased, ultimately resulting in enhanced yield by approximately 16.67% [33].

Considering the above-mentioned facts and the existing evidence in literature that plasma-activated water has eco-friendly, yield-boosting, antifungal, antibacterial, and seed germination enhancement properties, as well as the potential to improve plant the defence mechanisms and resistance against environmental stresses, we applied PAW to Common beans, which are very important food plants all over the world.

This study is aimed at (A) testing seed priming with PAW as a novel approach for seed treatment to increase growth parameters, (B) testing the effect of seed priming with PAW on the growth and development of Common beans, (C) assessing physiological responses of Common beans to the seed priming combined with PAW treatments.

The first part of our research was conducted as laboratory-based screening of various PAW treatments of seed germination parameters to identify the most effective PAW treatment. Subsequently, the selected PAW treatment was applied under outdoor conditions to assess its potential for enhancing the plant growth. The experimental approaches encompassed direct application of PAW through seed priming, foliar spraying, and a combination of both methods in outdoor settings. Many parameters were obtained during the outdoor experiment during which bean plants grew to the physiological maturity. Parameters, such as seedling emergence, plant height, plant biomass parameters, and selected physiological parameters well explained and described the effect of seed priming with different uses of PAW. This study is one of the first that combines both very effective methods of seed priming of PAW and the foliar spray of PAW on Common beans in outdoor conditions.

## 2. Experimental Section

### 2.1 Transient Spark (TS) Discharge

PAW was prepared by batch treatment of regular tap water (pH 7.9) with multiple parallel transient spark (TS) discharges, typically 170 mL of tap water was treated by 17 discharges during 5- or 10-minutes. Schematic diagram of the TS discharge setup is shown in Figure 1A. The vessel dimensions are 28×17.5×1 cm. The distance between nozzles in the same row is 4 cm, and the distance between adjacent rows is 3.5 cm. Each TS was set at approximately 1 and 1.5 kHz repetition frequency of the pulses and 1 cm distance of the high voltage needle electrodes from the water surface. We used stainless steel needle electrodes and there was no observable electrode erosion during the water treatment. The PAW was utilized on the same day it was prepared. Voltage and current-voltage waveform of the TS discharge are shown in Figures 1B and 1C respectively.

The TS plasma power was quantified by the measurement of voltage, current, and frequency. The instantaneous power, P(t), at any time t, is defined as:

$$P(t) = V(t).I(t) \qquad (1)$$

Where *V(t)* is the voltage, and *I(t)* is the current at every time *t*. To find the average power over one period, we integrate *P(t)* over the period (*T*) and divide by *T*, i.e. multiply by the frequency (*f*):

$$P_{avg} = \frac{1}{T}\int_0^T P(t)dt = f\int_0^T V(t).I(t)\,dt \qquad (2)$$

The typical average power of the multipin TS is 42.5 W, i.e. 2.5 W per needle for the frequency of 1 kHz, and 54.4 W, i.e. 3.2 W per needle for *f*=1.5 kHz.

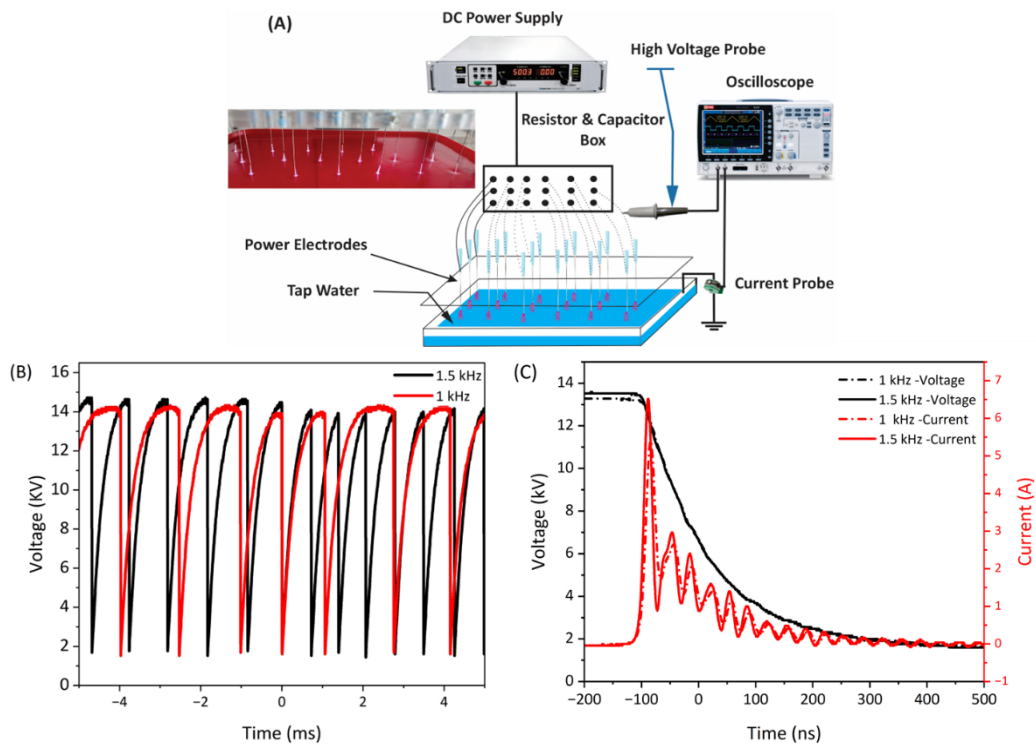

FIGURE 1 (A) schematic of multi-pin transient spark discharge set-up, (B) voltage waveforms, (C) current-voltage waveform at 1 and 1.5 kHz.

**2.2 Fountain Dielectric Barrier Discharge (FDBD)**

Figure 2A shows the schematic diagram of the FDBD reactor and water circulating system. An AC neon transformer (15 kV peak to peak, 20 kHz) and a variable autotransformer were used for generating the plasma. A DC power supply was used to control the water pump flow rate. The reactor is suitable for the treatment of relatively large amount of water with 85 mL/min flow rate. The reactor is a coaxial cylinder, and the dielectric used was a glass tube 180 × 25 mm in size and 1.5 mm thick. The central electrode is copper, and at the other electrode is a copper coil wrapped around the dielectric. The spacing between the tubes (discharge gap) is 3 *mm*. After being fed into the reactor, water flows laminarly up through the central electrode tube zone before dropping out of its top end and falling down on its exterior, thus making a thin water layer exposed to the DBD micro-discharges (120 mm discharge zone). One litre of water was treated for 20 and 30 minutes. Figure 2B and 2C show the voltage and the current-voltage waveform of the FDBD, respectively.

We used the Lissajous figure method to quantify the DBD dissipated power. A reference capacitor ($C_m$ = 6 nF) was connected in series with the FDBD device to measure the charge transferred during the plasma discharge. The voltage across the FDBD electrodes and the measuring capacitor was simultaneously recorded using high voltage probes and a digital oscilloscope. The charge $Q$ was calculated from the voltage across the capacitor using the relation $Q = C_m \cdot V_C$. A Lissajous figure was generated by plotting DBD voltage on the x-axis against the estimated charge on the y-axis throughout one complete AC cycle. The area covered by the resulting figure represents the energy dissipated every cycle (Figure 2C), corresponding to a discharge power of 48.8 W [34].

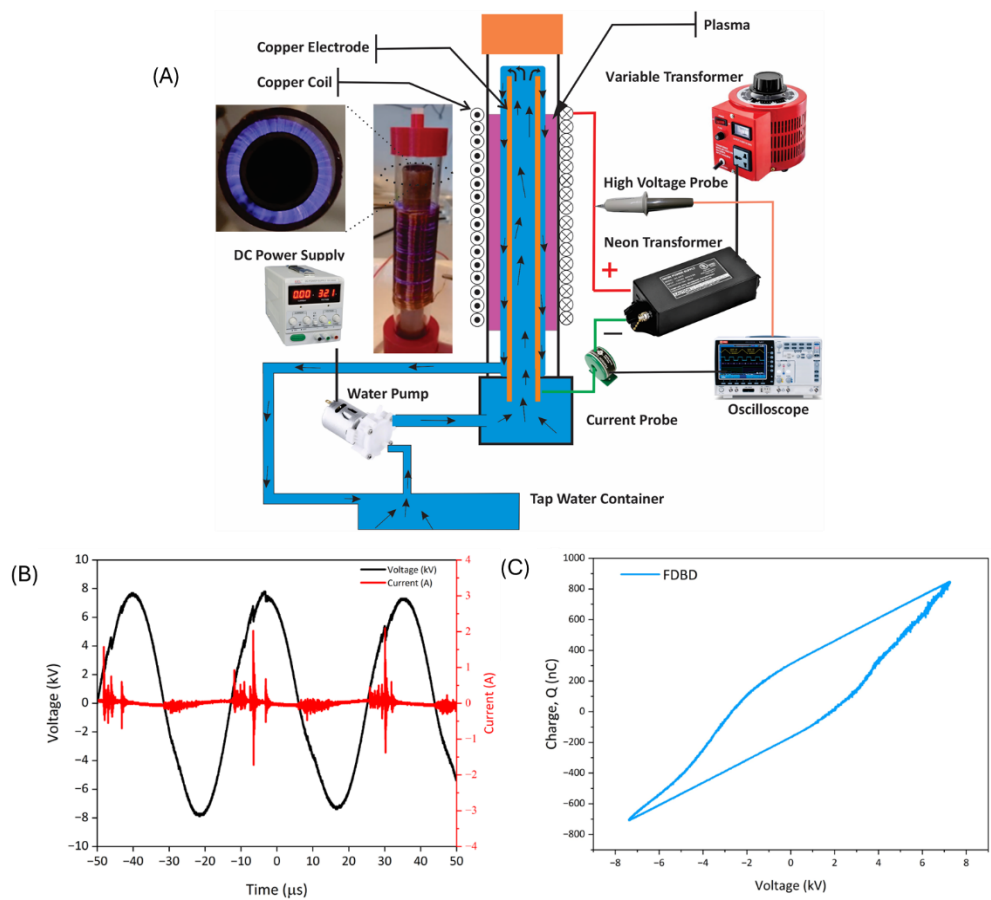

FIGURE 2 Fountain Dielectric Barrier Discharge, (A) schematic diagram and real picture, (B) current-voltage waveform, and (C) Lissajous voltage-charge figure.

## 2.3 Microwave Plasma (MW)

The water activated using MW plasma was prepared by our industrial partner (Proline solutions, Austria). The microwave power supply operates at a frequency of 2.45 GHz and the MW plasma jet is directed towards the treated water surface. Ambient air was used as a working gas. All PAW parameters were measured after plasma treatment and before its application to plants and seeds. A schematic diagram of MW plasma is shown in Figure 3. The plasma system utilizes 1.8 kW of electrical power and generates 45 L of PAW per hour. More detailed specifications on this MW plasma activated water are detailed in reference [35].

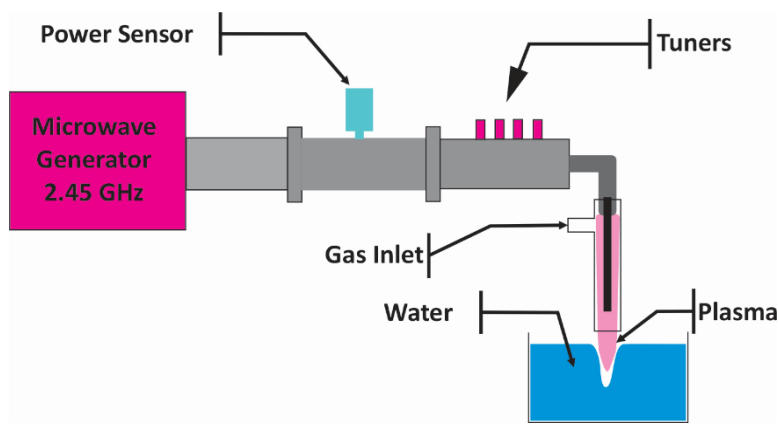

FIGURE 3 Schematic of Microwave Plasma Jet set up in Proline solutions (Austria).

## 2.4 Plasma-Activated Water Parameters

The experimental discharge set-ups are shown in Figures 1-3 (Multi-Pin Transient Spark, Fountain Dielectric Barrier Discharge, and Microwave Plasma Jet). In TS and FDBD set-ups, a high voltage probe (Tektronix P6015A) was used for voltage measurement. A Rogowski current meter (Pearson Electronics 2877) measured the discharge current. The Tektronix TDS 2024 digitizing oscilloscope recorded and analysed the temporal evolution of electrical discharge parameters (voltage and current). The PAW was produced from tap water available in Bratislava, Slovakia, with ~0.5 mS/cm conductivity. Unlike deionized or distilled water, it is more physiologically appropriate for plants and has a natural buffering ability to maintain a steady pH due to its calcium-carbonate buffer system. Consequently, the PAW produced from the tap water after TS and FDBD discharge treatment/activation does not significantly modify its original pH.

### 2.4.1 Physiochemical Parameter of PAW

The pH was measured using a portable meter (WTW 3110, Weilheim, Germany). Electrical conductivity and temperature were measured with a digital conductivity meter (GMH 3430, GREISINGER electronic, Germany). TDS was measured by 4-in-1 Multi Meter (Noyafa Digital, Philippines).

### 2.4.2 Hydrogen Peroxide and Ozone Measurement in PAW

The quantity of dissolved $H_2O_2$ was measured by its interaction with titanyl ions from titanium oxysulfate $TiOSO_4$ reagent, yielding a yellow product of pertitanic acid with a maximum absorbance peak at 407 nm:

$$Ti^{4+} + H_2O_2 + 2H_2O \rightarrow TiO_2H_2O_2 + 4H^+ \qquad (3)$$

UV-VIS spectrometer (Shimadzu 1900) was used for detecting the absorption spectra. The yellow-coloured product generated is stable for a minimum of 6 hours. To counteract the presence of nitrites in the PAW and prevent a potential decay of plasma-treated water under acidic conditions, a 60-mM solution of sodium azide ($NaN_3$) was added to the samples containing $H_2O_2$ prior to their combination with the reagent ($TiOSO_4$) to prevent the decomposition of $H_2O_2$ by nitrites. Sodium azide promptly converts nitrites into molecular nitrogen under acidic conditions.

$$3N_3^- + NO_2^- + 4H^+ \rightarrow 5N_2 + 2H_2O \qquad (4)$$

After the combination of the samples with NaN3, we added the titanium oxysulfate reagent into the sample. The $TiOSO_4$ reagent to sample ratio is 2:1.

The indigo blue colorimetric method was used for ozone ($O_3$) concentration measurement. In acidic solution, ozone rapidly decolourizes indigo blue. The decrease in absorbance is linear with increasing concentration. The maximum absorbance peak was measured at 600 nm.

### 2.4.3 Nitrite and Nitrate Measurement

The concentration of $NO_2^-$ was measured after calibration by using a commercial kit containing Griess reagent (Cayman Chemicals, Ann Arbour, MI, USA), resulting in a pink azo-product with a maximal absorbance peak at 540 nm [36].

For $NO_3^-$ measurement we used 2,6- dimethylphenol (DMP) (Sigma-Aldrich, USA) reagent. In the sulfuric and phosphoric acid solution nitrate it forms 4-nitro-2,6-dimethylphenol that is then detected photometrically at ~330 nm from the contribution of both $NO_2^-$ and $NO_3^-$, then by subtracting the $NO_2^-$ concentration, we measured $NO_3^-$ [37].

Specifically, the detection limits for the methods were as follows: hydrogen peroxide ($H_2O_2$) – 1 µM, ozone ($O_3$) – 0.5 µM, nitrite ($NO_2^-$) – 1 µM, and nitrate ($NO_3^-$) – 2 µM. These limits are within the sensitivity range of the respective spectrophotometric assays and are consistent with previous validation studies [36, 37].

Possible interferences among reactive species were carefully minimized during measurements. For $H_2O_2$ quantification, sodium azide ($NaN_3$) was added to eliminate the effect of nitrites, which can otherwise decompose $H_2O_2$ under acidic conditions. The indigo blue method used for $O_3$ detection is highly specific, as ozone rapidly decolorizes indigo dye without significant interference from other long-lived RONS such as $H_2O_2$ or $NO_X$ species. Nitrite and nitrate concentrations were determined using Griess reagents and the 2,6-dimethylphenol method, respectively, both of which are selective for their target analytes when performed under controlled acidic conditions.

**2.5 Seed Samples and Seed Priming**

Seeds of *Phaseolus vulgaris* L. were obtained from the Crop Research Institute in Prague, Czech Republic. The seeds were stored in standard dry conditions prior to use. Only ripe, intact, surface-untreated (unpeeled), and surface-correct seeds were used in the experiment [38].

Seed priming was performed by immersing 150 seeds in 100 mL of the PAW solution (one type of discharge treatment) for 6 hours. The soaking process was carried out in glass flasks covered with aluminum foil. Then the seeds were taken out and slowly dried at room temperature (22 °C) for 48 hours. In this way, the seeds were treated with the 3 different types of PAWs prepared under different conditions by TS, FDBD, and MW discharge set-ups (Table 1). The design of the laboratory experiment was supplemented with a control set with untreated seeds (K) and a control set with seeds that were primed with only tap water (TW).

TABLE 1 Plasma discharge systems and their parameters for PAW preparation.

| Plasma | Voltage (kV) | Frequency (kHz) | Power (W) | Time (minutes) | PAW (mark) |
|---|---|---|---|---|---|
| - | - | - | - | - | K |
| - | - | - | - | - | TW |
| TS | 13.6 | 1.50 | 54.4 | 5 | A |
| TS | 13.6 | 1.50 | 54.4 | 10 | B |
| TS | 13.2 | 1.00 | 42.5 | 5 | C |
| TS | 13.2 | 1.00 | 42.5 | 10 | D |
| FDBD | 15.0 | 20.00 | 48.8 | 20 | E |
| FDBD | 15.0 | 20.00 | 48.8 | 30 | F |
| MW | - | 2.45 | 1800.0 | - | G |

## 2.6 Laboratory Seed Experiments

Pre-treated and control seeds were placed in a zigzag pattern (to ensure proper spacing) on moist laboratory-grade filter paper, which was then rolled into a tower. We used 30 seeds per 1 filter paper and 5 repetition for each treatment. The filter papers for one treatment were sealed in plastic bags, placed upright in a culture box, and incubated in the dark at 25 °C.

Numbers of germinated seeds were recorded every 12 hours intervals throughout the experiment. The seed germination (SG, %) was calculated as a number of germinated seeds divided by their total number (30). A seed was considered to have germinated if a radicle appeared and was 2 mm long [38]. For the outdoor experiment, the most effective type of PAW (marked B, TS-1.5kHz 10m, see Table 1) from the laboratory experiment was selected, according to the seed germination.

## 2.7 Outdoor Pot Trials

Outdoor planting experiment was conducted in summer 2024 at the Faculty of Natural Sciences, Comenius University Bratislava, located at coordinates 48°08'57.6"N latitude and 17°04'19.9"E longitude. During the experiment, the basal weather conditions were recorded at a nearby local meteorological station maintained by the Faculty of Mathematics, Physics and Informatics, Comenius University Bratislava (FMFI CUB), located at coordinates 48°09'03.9"N latitude and 17°04'07.0"E longitude (Figure 4).

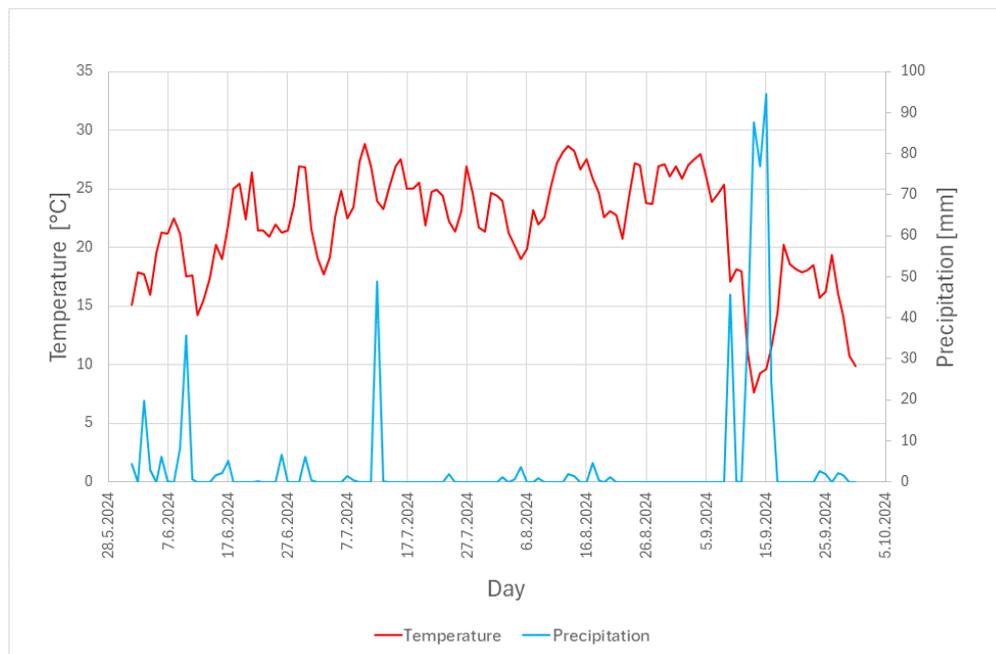

FIGURE 4 Average daily temperature (Temperature) and total daily precipitation (Precipitation) measured by the meteorological station at FMFI CUB during outdoor plant cultivation.

In the outdoor experiment, only PAW made by TS discharge operating at 1.5 kHz and treating water for 10 min (marked B, see Table 1). The plant treatments were designed into 4 types: control (seed priming with tap water), priming (seed priming with PAW), spraying (foliar application of PAW) and priming*spraying (combination of both previous treatments). The seed priming treatment was done the same way as in the laboratory experiment. The PAW as foliar spray was initiated on day 12, when seedlings were in the three leaved stage, and was sprayed on days 12, 19, 25, 32, and 39, subsequently. A high-pressure nozzle bottle was used to ensure the even distribution of the spray on the leaves. Each plant was sprayed ~3ml of PAW during every application of foliar spray.

Plastic pots (90 L) filled with commercial peat soil (Substrát univerzálny záhradný, fa. Nature Garden, Slovakia) were used for the outdoor experiments (Figure 5). Each treatment consisted of three replications and the pots were spatially arranged using a completely randomized design (CRD) in outdoor conditions. Three bean seeds were placed in each of 9 holes (in clip of 3 × 3 holes, each 3-4 cm deep), i.e. a total of 27 seeds were sown in one pot (1 pot ~ 1 replication, 1 treatment ~ 81 seeds).

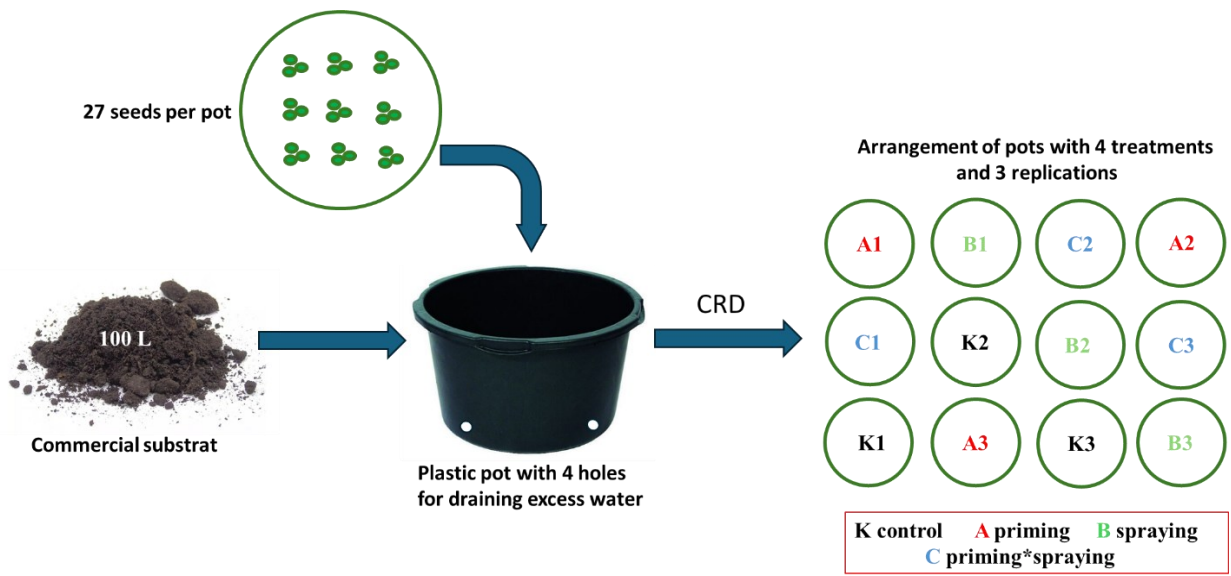

FIGURE 5: Design of the outdoor experiment. CRD completely randomized design.

### 2.7.1 Seedling Emergence, Shoot Length and Plant Biomass Parameters

The pots were checked every day and, if necessary, the soil was watered (always the same amount for each pot). The number of growing seeds was monitored within 7 days from the initial sowing. The shoot length was measured every day (after 24 hours) up to 64 days after sowing. The experiment was carried out for 64 days because of the severe weather conditions that affected several central European countries in mid-September 2024. Nevertheless, this duration does not encompass the entire life cycle of common beans. At the end of the experiment, all plants were harvested and divided into underground (roots) and above-ground parts (shoot). The fresh biomass was weighed (scale sensitivity 0.01 g) and then dried in a thermostat at a temperature of 50 °C for 48 hours. Seed emergence (%) was determined as the ratio of growing seeds to the number of sown seeds [38]. Shoot length (cm) was determined as the average of the tallest plants from 9 holes for each pot (replicates). The weight of fresh root and shoot (g), as well as the weight of dried root and shoot (g), were determined as the average of the harvested biomass for each pot (pot = replicate).

R/S fresh weight was calculated as the ratio between the weight of fresh biomass of the underground part to the fresh weight of the above-ground part of all plants from one pot; R/S dried weight was determined similarly. Parameters of vitality index II. (g) and vitality index III. (g) were determined according to the following formulas:

vitality index II. (g) = seed emergence (%) × fresh weight of all plants from one pot (g)     (5)

vitality index II. (g) = seed emergence (%) × dry weight of all plants from one pot (g)     (6)

All these measured and calculated parameters were determined according to the methodological approaches in [38].

**2.7.2 Protein Extraction and Determination**

Ten young leaves of beans plants (counted from the shoot apex) were collected 64 days after sowing for protein determination and analysis of selected antioxidant enzymes activities. All samples were immediately frozen and stored in a in the freezer at a temperature of -16 °C. Approximately 1–1.5 g of young leaves were grinded in liquid nitrogen with a mortar and pestle. The homogeneous powder was suspended in 50 mM Na-phosphate buffer (pH 7.8) containing 1 mM EDTA, polyvinylpolypyrrolidone (1 g per 50 ml), protease inhibitor (Protease Inhibitor Cocktail) and centrifuged at 12,000 g for 30 min at 4 °C. The supernatant was used for the measurements of soluble protein concentrations, which were determined according to Bradford (1976) at $\lambda = 595$ nm, and for selected antioxidant enzymes (superoxide dismutase, guaiacol peroxidase, catalase, ascorbate peroxidase and glutathione reductase) [39].

Glutathione reductase (GR) activity was determined spectrophotometrically according to Foyer and Halliwell (1976) based on oxidation of NADPH at $\lambda = 340$ nm. The activity of GR was calculated with the molar extinction coefficient of NADPH 6.2 according to Claiborne (1985) and expressed as the decrease in NADPH in $\mu mol \cdot min^{-1} \cdot mg^{-1}$ of soluble proteins [40].

Guaiacol peroxidase (G-POX) activity was determined spectrophotometrically according to Frič and Fuchs (1970) based on the oxidation of guaiacol to tetraguaiacol in the presence of $H_2O_2$ at $\lambda = 440$ nm. The activity of G-POX was calculated with the molar extinction coefficient of tetraguaiacol 26.6 according to Claiborne (1985) and expressed in μmol of tetraguaiacol $\cdot min^{-1} \cdot mg^{-1}$ of soluble proteins [41, 42].

Catalase (CAT) activity was determined according to Hodges et al. (1997) based on $H_2O_2$ decomposition by catalase at $\lambda = 240$ nm. The activity of CAT was calculated with the molar extinction coefficient of $H_2O_2$ 39.1 according to Claiborne (1985) and expressed in nmol of $H_2O_2 \cdot min^{-1} mg^{-1}$ of soluble proteins [43, 44].

Ascorbate peroxidase (APX) activity was determined spectrophotometrically according to Nakano and Asada (1981) methods based on the oxidation of ascorbate at $\lambda = 290$ nm [45]. The activity of APX was calculated with the molar extinction coefficient of ascorbate 2.8 according to [41] and expressed by a decrease of ascorbate in μmol $min^{-1} mg^{-1}$ of soluble proteins.

Superoxide dismutase (SOD) activity was measured and expressed according to [39] by following the inhibition of the photo-reduction of 3- (4,5- dimethyl-2-thiazolyl)-2,5-diphenyl-2H-tetrazolium bromide (MTT assay) to the purple product formazan. Reaction was initiated by placing the reacting mixture in tubes and exposing it to cold-white light (50 $\mu M\ m^{-2}\ s^{-1}$) for

20 min. Non-illuminated and illuminated reactions without supernatant served as calibration standards. Reaction products were measured spectrophotometrically at $\lambda = 560$ nm. One unit of SOD activity was defined as the amount of protein necessary to inhibit 50% of the initial MTT reduction under light exposure. The resulting activity was then normalized to 1 mg of protein in the sample.

### 2.7.3 Determination of Chlorophyll Concentration

For determination of assimilation pigments, we followed the protocol of [46]. Young bean leaves were thoroughly ground with a mixture of magnesium carbonate and sea sand to achieve a homogeneous mixture and prevent the formation of pheophytin. Pigments were then extracted using 80% acetone, and the samples were centrifuged at 5,000 g for 5 minutes at 4 °C. The chlorophyll and carotenoid contents in the supernatant were quantified spectrophotometrically at 663.2 nm (chlorophyll *a*), 646.8 nm (chlorophyll *b*), and 470 nm (carotenoids).

### 2.7.4 Determination of Soluble Phenolics

The content of soluble phenolics was determined in young leaves according to the protocol of [47]. As a basis, phenolic content was measured using Folin-Ciocalteu reagent. The results were derived from a calibration curve of gallic acid (0 – 500 µg/mL) and expressed in gallic acid equivalents (GAE) per milligram of the sample fresh weight. The absorbance was measured at a wavelength of $\lambda = 765$ nm.

### 2.8 Data Analyses

The data obtained were analysed using the Statistica software [48] at a significance level of 0.05. One-way analysis of variance (ANOVA) and post hoc comparison (PHC) were performed. For PHC, Newman-Keuls test was used to evaluate the differences among the treatments in relation to bean development, growth, biomass and physiological parameters. The dependent variables were measured parameters, while the independent categorical variable were the treatments (control, priming, spraying, priming*spraying).

## 3. Results

### 3.1 Parameter of PAW

Temperature and pH of used PAW are shown in FIGURE A. Although there are increase of temperature in some PAWs, we kept them in the lab for 10 minutes to reach the room temperature like normal tap water (control) used. The pH for tap water (control) was 7.9. There was no big change in pH for TS groups and FDBD at these treatment times 5 and 10 min for TS, and 20 and 30 min for FDBD. The pH measured for water prepared by microwave plasma was substantially reduced to the lowest pH=4.

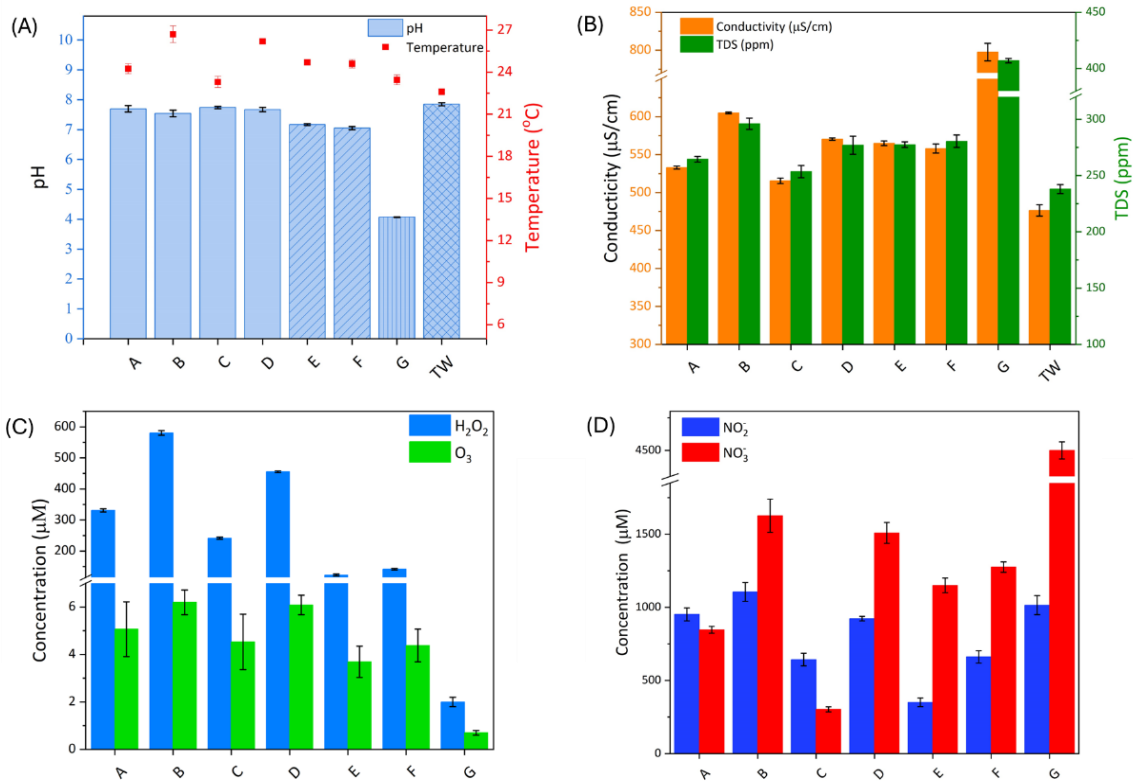

FIGURE 6 PAWs parameters after preparation. (A) pH and temperature, (B) conductivity

(orange) and TDS (green), (C) concentration of $H_2O_2$ (blue) and $O_3$ (green), (D) concentration of $NO_2^-$ (blue) and $NO_3^-$ (red). The PAWs designation used is shown in Table 1.

Figure 6B illustrates the conductivity (µS/cm) and the total dissolved solids (TDS, ppm) in PAW across various plasma treatment parameters. MW plasma treatment shows the highest conductivity (797.5 µS/cm) and TDS (407 ppm). Tap water control (TAP Water Cont) has the lowest conductivity (476.5 µS/cm) and TDS (238 ppm). TS-1.5 kHz and TS-1 kHz treatments lead to moderate increases in conductivity and TDS. FDBD treatments also increase conductivity and TDS, but slightly less than TS treatments.

Figure 6C represents the concentrations of hydrogen peroxide ($H_2O_2$) and ozone ($O_3$) in PAW under different experimental conditions. $H_2O_2$ concentrations (blue bars) are significantly higher than $O_3$ (green bars) in all conditions. The highest $H_2O_2$ concentration is observed at TS-1.5 kHz 10 min (580 µM). The lowest $H_2O_2$ concentration (excluding MW) is found in FDBD-20min (122 µM). $O_3$ concentrations remain relatively low, in coherence with very low $O_3$ solubility in water, with values ranging between 4–7 µM, showing less variation across conditions. The TS 1.5 kHz frequency consistently resulted in higher $H_2O_2$ concentrations than 1 kHz at the same treatment duration, indicating that a higher frequency enhances $H_2O_2$ generation. Longer treatment times (10 min) lead to more $H_2O_2$ production than shorter times (5 min) for both 1.5 kHz and 1 kHz conditions. For $O_3$, the increase is not significant with longer exposure time, suggesting a saturation effect. FDBD treatments (20 min and 30 min) produce lower $H_2O_2$ concentrations than TS treatments. MW treatment shows the lowest levels of both $H_2O_2$ (2 µM) and $O_3$ (0.2 µM) in PAW.

Figure 6D shows the concentrations of nitrogen oxides ($NO_2^-$ and $NO_3^-$) under different plasma treatment conditions. $NO_3^-$ (red bars) is consistently higher than $NO_2^-$ (blue bars) in all conditions. MW PAW had the highest $NO_3^-$ and $NO_2^-$ concentrations, with $NO_3^-$ exceeding

4500 µM. TS-1.5 kHz 10 min produces the highest NO$_3^-$ among all plasma-treated samples (~1700 µM). FDBD treatments resulted in moderate NO$_3^-$ and NO$_2^-$ concentrations, lower than TS-1.5 kHz 10 min but higher than TS-1 kHz 5 min. TS-1.5 kHz produces more NO$_3^-$ and NO$_2^-$ than TS-1 kHz, suggesting that a higher frequency enhances nitrogen oxide formation, similar to H$_2$O$_2$. TS-1.5 kHz 10 min produced very high NO$_3^-$ concentration (~1700 µM), while TS-1 kHz 10m is slightly lower (1508 µM). NO$_2^-$ concentrations show a similar trend but remain lower than NO$_3^-$.

### 3.2 Laboratory Seed Experiments

An overview of bean seed germination parameters after seed priming with three PAWs, tested under controlled conditions, is provided in Figure 7. It was observed that all treated seeds exhibited lower germination compared to the tap water control. This finding is noteworthy, as it indicates that the plasma-activated water treatments influenced the early germination behavior of seeds. Among the treatments, the highest values within the treated sets were recorded in treatment B (TS- 1.5 kHz, 10 min; see Table 1), which also produced visually healthier seedlings despite the overall reduction in seed germination. Based on these observations, treatment B was selected for subsequent outdoor experiments to further assess its effects under outdoor field conditions.

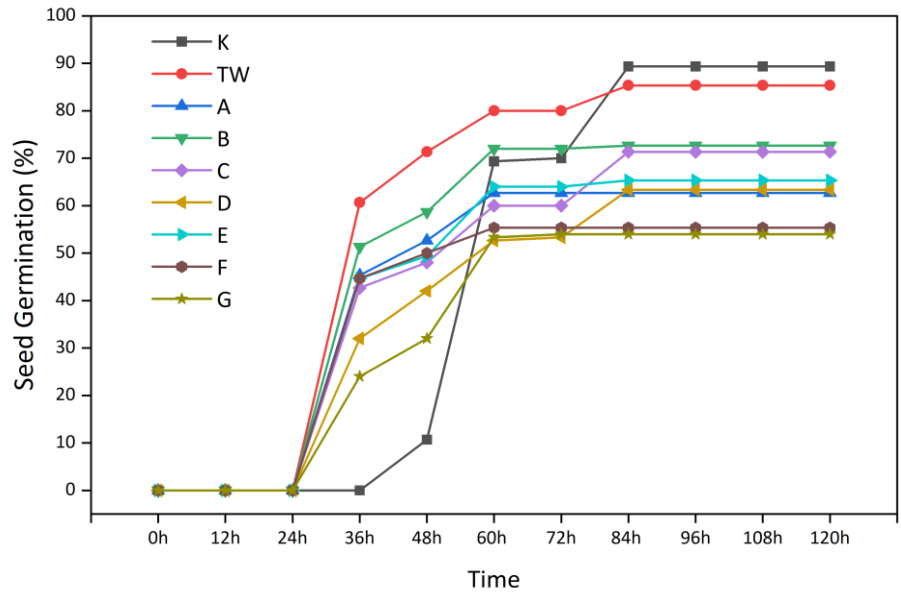

FIGURE 7 Seed germination (%) of *Phaseolus vulgaris* after seed priming with different PAWs (for more detail see methodology and Table 1).

### 3.3 Outdoor Pot Trials

The first seedlings appeared on the 4[th] day after sowing. No significant difference was found in seedling emergence between treatments, the highest value was recorded in seeds that underwent priming (76.33%, Table 2).

TABLE 2 Seedling emergence, shoot length and plant biomass parameters in *Phaseolus vulgaris* after various PAW treatments measured the last day of the experiment. Different letters represent significant differences among the treatments at P < 0.05; the highest measured values are marked in bold, (for more details see methodology).

| Treatment | Seedling emergence (%) | Plant height (cm) | Weight of fresh shoot (g) | Weight of dried shoot (g) | Weight of fresh root (g) |
|---|---|---|---|---|---|
| Control | 66.67 ± 5.78$^a$ | 128.78 ± 5.44$^a$ | 566.33 ± 119.84$^a$ | 178.33 ± 8.19$^a$ | 65.66 ± 3.28$^a$ |
| Priming | **76.33 ± 3.28$^a$** | **173.63 ± 3.46$^b$** | **828.33 ± 171.50$^a$** | **271.00 ± 31.26$^b$** | **101.66 ± 1.45$^b$** |
| Spraying | 69.00 ± 1.00$^a$ | 139.55 ± 1.64$^a$ | 560.66 ± 137.42$^a$ | 210.00 ± 15.82$^{ab}$ | 92.66 ± 6.44$^b$ |
| Priming*Spraying | 71.33 ± 1.33$^a$ | 159.70 ± 9.67$^b$ | 749.33 ± 40.34$^a$ | 220.00 ± 7.09$^{ab}$ | 89.00 ± 2.65$^b$ |

| Treatment | Weight of dried root (g) | R/S_fresh weight | R/S_dried weight | Vitality index II. (kg) | Vitality index III. (kg) |
|---|---|---|---|---|---|
| Control | 20.66 ± 1.20$^a$ | 0.13 ± 0.02$^a$ | 0.12 ± 0.01$^a$ | 43.51 ± 12.22$^a$ | 13.35 ± 1.67$^a$ |
| Priming | **36.33 ± 1.20$^c$** | 0.13 ± 0.03$^a$ | 0.14 ± 0.01$^a$ | **72.10 ± 15.69$^a$** | **23.67 ± 3.40$^b$** |
| Spraying | 32.00 ± 1.53$^b$ | **0.18 ± 0.04$^a$** | **0.15 ± 0.01$^a$** | 44.85 ± 9.35$^a$ | 16.67 ± 0.93$^{ab}$ |
| Priming*Spraying | 29.33 ± 0.88$^b$ | 0.12 ± 0.01$^a$ | 0.13 ± 0.01$^a$ | 59.71 ± 17.93$^a$ | 17.78 ± 0.41$^{ab}$ |

The growth of the plants (shoot length) throughout the whole duration of the outdoor experiment (64 days) is shown in Figure 8. From the beginning of the experiment, the tallest plants were treated with seed PAW priming. Conversely, the shortest plants were those treated with continuous PAW spraying and with control set (Figure 8). At the end of the experiment, a significant difference (P < 0.05) was found between the individual treatments: the priming treatment and the priming*spraying combination were significantly higher from both spraying and control sets (Table 2).

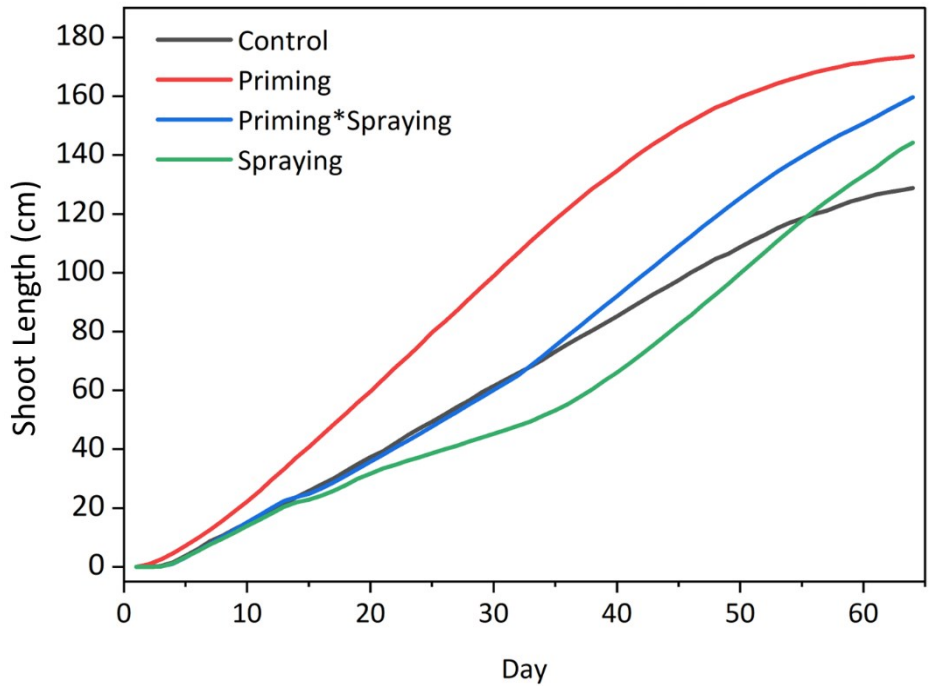

FIGURE 8 Kinetic increase in shoot length of *Phaseolus vulgaris* after different PAW treatments (measured every day). Measured from the first day of seed emergence to the 64th day when the experiment was finished.

Regarding the parameters related to the plant biomass growth, a significant difference ($P < 0.05$) was found in four measured parameters: weight of dry shoot, weights of fresh and dry roots, and vitality index III (Table 2). For these listed parameters, the highest values were always recorded for the sets of plants with seed priming treatment. The highest value for weight of dried shoot was 271 g, which is 152% compared to the control. The highest values for weights of fresh and dried roots were 102 g and 36 g, which corresponds to 155% and 176% respectively (relative to the control). The highest value for vitality index III was 24 g, which is 177% compared to the control set (Table 2).

Physiological parameters for which a significant difference was found between individual treatments are presented in Figure 9, those parameters where the difference was not found are presented in Table 3. In all studied PAW treatments, SOD activity was found higher than in the control, with the most significant increase observed in the priming treatment (Figure 9E), but with no differences between the spraying and priming*spraying treatments. G-POX activity exhibited a significant increase in spraying treatment (Figure 9B) in comparison to all other treatments. We also observed a significant decrease in priming*spraying treatment compared to all other treatments monitored. CAT activity increased significantly across all PAW treatments in comparison to the control, with the highest increase recorded in the priming*spraying treatment (Figure 9C). Similarly to SOD, APX and GR activities showed the highest increase in the priming treatment set compared to the control set (Figure 9E, 9D and 9A).

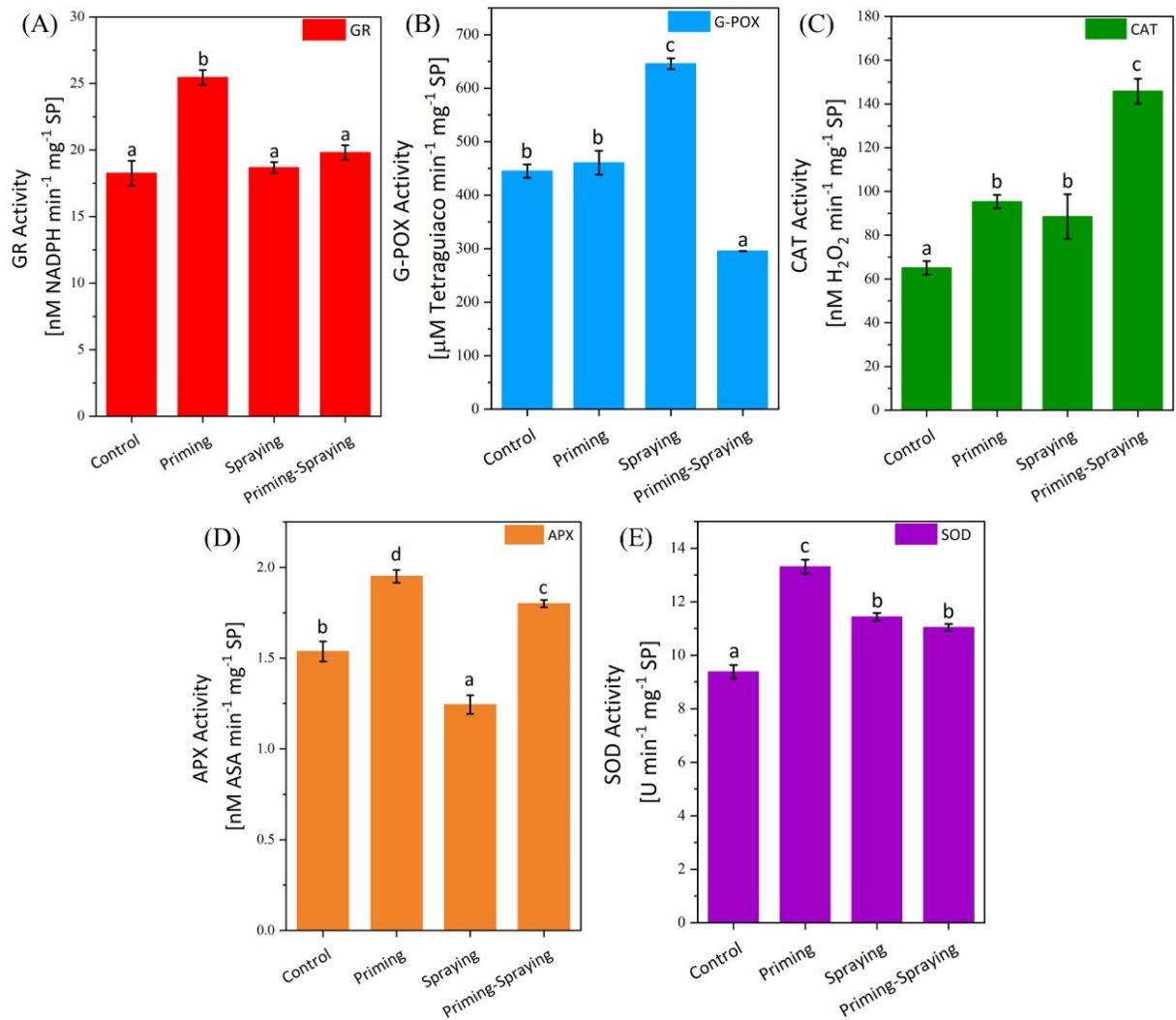

FIGURE 9 Activity of GR (A), G-POX (B), CAT (C), APX (D), SOD (E), the antioxidant enzymes, from young leaves of *Phaseolus vulgaris* under control and PAW treatments. Different letters represent significant differences among the treatments at P < 0.05 according to the Newman-Keulsuv test (number of replicates = 3).

TABLE 3 Concentrations of chlorophylls a, b, and carotenoids in mg g$^{-1}$ of fresh weight and total content of soluble phenols (GAE – Galic Acid Equivalents) in young leaves of *Phaseolus vulgaris* after different PAW treatments. Different letters represent significant differences among the treatments at P < 0.05 (number of replicates = 3, the highest measured values are marked in bold).

| Treatment | Chlorophyll a (mg.g$^{-1}$) | Chlorophyll b (mg.g$^{-1}$) | Carotenoids (mg.g$^{-1}$) | Phenols (GAE) |
|---|---|---|---|---|
| Control | 0.72 ± 0.07$^a$ | **0.35 ± 0.08$^a$** | 0.15 ± 0.01$^a$ | 2.61 ± 0.36$^a$ |
| Priming | 0.82 ± 0.08$^a$ | 0.31 ± 0.04$^a$ | 0.17 ± 0.03$^a$ | 2.16 ± 0.19$^a$ |
| Spraying | 0.71 ± 0.05$^a$ | 0.27 ± 0.02$^a$ | **0.18 ± 0.01$^a$** | **2.81 ± 0.31$^a$** |
| Priming*Spraying | **0.83 ± 0.10$^a$** | 0.32 ± 0.03$^a$ | **0.18 ± 0.02$^a$** | 2.73 ± 0.19$^a$ |

Table 3 represents the concentrations of chlorophyll *a*, chlorophyll *b*, carotenoids, and phenolic compounds. No significant differences were observed in any of these parameters across the individual treatments.

## 4. Discussion

### 4.1 Parameter of PAW

The three PAWs used in this study exhibited markedly different physicochemical characteristics, reflecting the nature of their discharge systems. TS discharge, operating at atmospheric pressure with repetitive high-voltage nanosecond pulses, produced the highest concentration of both hydrogen peroxide and nitrate (up to 580 µM and 1700 µM, respectively). This is attributed to the strong transient electric fields and higher electron energies that promote gas-phase dissociation of oxygen and nitrogen molecules, resulting in efficient transfer of RONS into the liquid phase. FDBD, by contrast, generated moderate RONS levels with lower oxidative potential. Its continuous AC microdischarges create a more uniform plasma with lower energy density, leading to milder chemical activation of the liquid. MW discharge, while operating at much higher total power (1.8 kW), produced the lowest concentrations of long-lived RONS and showed a significant acidification (pH ≈ 4). This effect results from the high gas temperature and rapid recombination processes typical of microwave plasmas, which limit the dissolution of active species into water.

These differences in composition explain why the TS-generated PAW, particularly the TS-1.5 kHz 10 min treatment, was selected for further experiments. Its balanced mixture of oxidants ($H_2O_2$, $O_3$) and nitrogen oxides ($NO_2^-$, $NO_3^-$) likely provided an optimal redox and nutrient environment, stimulating bean seedling growth and antioxidant enzyme activation while avoiding the excessive acidity observed in MW PAW. Thus, the distinct discharge characteristics directly determined the chemical profile of each PAW, which in turn influenced plant physiological responses.

### 4.2 Laboratory Seed Experiments

PAW marked as Treatment B (TS–1.5 kHz for 10 minutes; see Table 1) was selected for the outdoor planting experiment based on its relative performance among the PAW treatments. Although the untreated control ultimately achieved the highest germination rate under laboratory conditions, Treatment B consistently showed the best results within the PAW-treated groups and produced seedlings that appeared more vigorous and healthier in early growth stages. This visual improvement, together with its intermediate germination performance, indicated potential physiological benefits worth testing under field conditions. As shown in Figure 7, both control groups (untreated seeds and those primed with tap water) outperformed

PAW B by the end of the laboratory test, confirming that conventional seed priming can effectively stimulate germination. Nevertheless, the selection of Treatment B was justified by its relatively superior performance among PAW variants and the observed early seedling vitality, which warranted further evaluation under outdoor conditions. Soaking seeds in PAW might not reveal notable differences in controlled germination tests, but it can enhance performance when planted in outdoor soil pots. This difference aligns with the prior research showing that the effects of priming are frequently reliant on the context of laboratory settings that might not entirely mimic field or soil conditions, where primed seeds gain advantages from improved stress resilience, nutrient absorption, and interactions with microorganisms [49]. Seed priming with PAW may enhance plant growth attributes overall as reported by several findings of many previous experiments focused on seed priming research, e.g. [50–54]. It was established that seed priming with PAWs can fasten and enhance germination by inducing bio-chemical changes in the seeds of different plant species. However, effect of PAW on germination varies based on plant species, discharge type, type of water used, (ionized, deionized), and experimental conditions i.e. *in vitro* vs. *in vivo* [27]. For instance, a study involving PAW treatments found no statistically significant differences in the germination rates among radish, tomato, and marigold plants when compared to the control groups when grown in pots [55]. On the other hand, another study reported a 50% increase in germination rate for lentil seeds subjected to the PAW treatment, compared to untreated seeds in *in vitro* conditions [21]. Furthermore, many studies have reported diverse effects of PAWs on seed germination, with some demonstrating positive influences, others showing no impact, and a few exhibiting negative effects. The synthesis of PAWs and seeds exposure time is critical in enhancing plant growth parameters [56]. Low levels of plasma RONS species may act as positive signaling stimuli to alleviate seed dormancy, but excessive quantities of these reactive species can accumulate and potentially become detrimental to seed germination [57].

The laboratory experiment served as an initial screening to determine the optimal PAW treatment, which was then further investigated in outdoor settings. As illustrated in Figure 7, the PAW treatments had a positive impact on plant growth parameters, exhibiting distinct effects outdoor experiment compared to the laboratory setting.

**4.3 Outdoor Pot Trials**

The outdoor pot experiment included seed PAW-priming and foliar spraying of seedlings and plants at different growth stages, as well as a combination of priming and spraying. Under outdoor conditions, it was significantly demonstrated that seed priming with PAW significantly improved shoot length and dry plant biomass production (Table 2, Figure 8). This result is coherent with several previous studies where plant height and biomass were significantly improved with respect to the control sets, when primed/sprayed with PAW [58–61]. This could be due to the PAW main constituents are ROS and RNS. Notably, hydrogen peroxide, nitrate, and nitrite are the measured long-lived reactive species within this group [62, 63]. The basic mechanism involved in plant germination could be that ROS may cause cracks in the seeds coats, making water and nutrient assimilation into seeds easier. Hence, the germination increased and gave rise to overall better plant growth and seedling lengths [64]. Our results indicate that PAW-priming did not enhance germination in laboratory settings, but in outdoor environments, these treatments resulted in enhanced emergence and seedling growth. This difference can be attributed to various physiological, biochemical, and ecological mechanisms, as supported by existing studies. When seeds take up water during priming, reactive oxygen species (ROS) can be produced [65]. At low to moderate levels, ROS function as signalling molecules that assist in overcoming seed dormancy, soften seed coats, modify hormone levels (such as decreasing abscisic acid and increasing gibberellins), and trigger the activation of antioxidant mechanisms. However, if ROS levels exceed the protective capacity of the seed, damage may occur, including membrane lipid peroxidation, electrolyte leakage, and DNA damage [61–63].

Additionally, mild oxidative stress experienced during priming has the potential to stimulate protective antioxidant mechanisms, thereby enhancing the seedlings' ability to cope with environmental changes [66]. Consequently, whereas the laboratory assessment revealed the short-term inhibitory impacts of PAW, the findings from outdoor conditions demonstrate the longer-lasting adaptive advantages of priming in natural settings, where chemical buffering and biological interactions promote the plant recovery and growth [65–67].

Foliar application puts the plants under stress in their initial phases of growth, however, with time the seedling regained better growth as compared to the control. This could be possibly due to excessive accumulation of ROS which were absorbed by leaves and hindered the plant growth. Similar phenomenon was observed when PAW was applied to green leafy vegetables [50]. Regarding the observed enhanced biomass production, PAW-inducing mechanisms for plant growth depend upon many factors to provide overall positive effect. Nitrate is the key constituent of PAW and as the best form of nitrogen that plants can assimilate, it contributes towards the acceleration of the plant growth [68]. Furthermore, nitrate nitrogen, when absorbed by plant cells, undergoes a stepwise reduction which is aided by the enzymes nitrate reductase and nitrite reductase. This process ultimately generates ammonium ions, which are then incorporated into amino acids, and could possibly promote the plant growth and enhanced seedling lengths [69].

Antioxidant enzymes play a crucial role in plant defence against environmental stresses by mitigating oxidative damage. SOD is a key component of this system, catalyzing the dismutation of superoxide radicals into $H_2O_2$, and thus represent the first line of defence against ROS [70]. To prevent oxidative damage, the resulting $H_2O_2$ is removed by a broader antioxidant enzyme network, including CAT, G-POX, APX, and GR. Catalase directly decomposes $H_2O_2$ into water and oxygen without the consumption of cellular reducing equivalents, providing an energy-efficient mechanism for oxidative protection [71]. In contrast, G-POX utilizes $H_2O_2$ as a substrate for other biochemical reactions, such as lignin polymerisation [72]. GR together with ascorbate peroxidase play a critical role in maintaining redox homeostasis by protecting cells against RONS and facilitating $H_2O_2$ detoxification through the ascorbate-glutathione cycle.

In our results, almost all treatments involving PAW applications, led to a significant increase in antioxidant enzyme (SOD, CAT, and APX) activities (Figure 9). In contrast, increases in G-POX and GR activities were observed only under specific application methods: G-POX activity significantly increased following foliar application, while GR activity was enhanced when seeds were primed with PAW. Similar plant responses have been reported by [73], who observed increased total antioxidant activity in pea seeds exposed to air plasma. Kostolani et al. [74] also documented significantly higher G-POX and CAT activity in barley and pea seeds watered with PAW. Consistent with our findings, Okruhlicová et al. [23] reported a significant increase in G-POX enzyme activity following foliar application across various maize hybrids. Likewise Ndiffo Yemeli et al. [75] reported enhanced CAT activity in barley and maize after PAW application, although they observed a significant decrease in SOD activity – an enzyme generally associated with the oxidative stress response. This observation differs from our findings, as SOD activity was significantly higher in all PAW treatments compared to the control group. Finally, many studies, including those by Kučerová et al. [22, 76] suggested that catalase is the dominant enzyme responsible for $H_2O_2$ scavenging in aboveground plant tissues, whereas G-POX predominates in root systems. Our results are consistent with these observations.

Increased activity of antioxidant enzymes in PAW treatments also correlates with the observed improved growth parameters. Since it is well established that enhanced antioxidant activity contributes to the improved plant growth and vitality [77], enhancing the antioxidative capacity represents one of the main adaptive responses in plants, not only to stress conditions [78]. In our study, we observed greater shoot growth in both priming and priming*spraying treatments. Moreover, increased biomass production was recorded not only in the shoots but also in the root system across all tested treatments, as discussed above. These results suggest that plants primed with PAW maintained better ROS homeostasis, which may have enabled them to grow more successfully than the controls [77]. The growth enhancement observed with PAW treatment seems to arise from a combination of enriched nutrients and redox-based signalling. The prominent nitrate signal identified in PAW indicates that a portion of the growth response may be due to extra nitrogen input, as nitrate is an easily absorbed nitrogen form that supports the synthesis of amino acids and proteins through the nitrate–nitrite–ammonium pathway mediated by nitrate and nitrite reductases [27, 79]. Concurrently, the steady increase in antioxidant enzyme activities (SOD, CAT, and APX) across PAW treatments suggests that low levels of RONS could function as signalling molecules, activating the plant's antioxidant defence system and improving their metabolic readiness for growth [77, 80]. The dual role of PAW, combining mild oxidative signalling and nutritional nitrogen input, has been highlighted in various studies that reported similar physiological responses in barley, maize, and pea treated with PAW [23]. The noticeable difference between laboratory and outdoor findings can be explained by the environmental buffering present in outdoor systems, where mild oxidative stress may lead to beneficial priming effects instead of inhibition [75]. Collectively, these results indicate that PAW functions not merely as a stress inducer but also as a biostimulant that provides both nutrients and signals. Future studies should incorporate nitrate-matched controls and biochemical assessments, such as nitrate reductase activity and total leaf nitrogen measurement, to more clearly delineate the nutritional and signalling roles in PAW-induced growth responses.

Assessing leaf pigment content holds considerable physiological importance, as chlorophylls (*a* and *b*) are essential for light energy utilization during photosynthesis. Chlorophyll concentrations directly influence photosynthetic efficiency and primary productivity. In addition, owing to their high nitrogen content, they also serve as indirect indicators of a plant's mineral nutritional status and overall metabolic activity [81]. Phenolic compounds are highly effective non-enzymatic scavengers of ROS due to their lower electron reduction potential comparing to the oxygen radicals. Additionally, phenoxyl radicals, which are formed when phenolic compounds neutralize oxygen radicals, are generally less reactive, thereby enabling phenolic compounds to quench ROS without promoting further oxidative reactions. Consequently, environmental stresses that induce oxidative stress often stimulate the synthesis of phenolic metabolites [47].

In our study, monitoring photosynthetic pigments and soluble phenolics in young bean leaves did not bring any significant differences across the various PAW treatments (Figure 9). Nitrate ($NO_3^-$) plays a crucial role in chlorophyll biosynthesis, as it serves as a primary nitrogen source for plants. Nitrogen, is a key structural element of the tetrapyrrole ring central to chlorophyll molecules. Thus, adequate nitrate availability typically promotes chlorophyll biosynthesis, enhancing photosynthetic capacity and plant growth [82]. Consistent with this, [81] demonstrated that seeds of *Lactuca sativa* L. treated with plasma-activated waters (PAWs) containing different $NO_3^-$ concentrations exhibited a significant increase in chlorophyll content under higher $NO_3^-$ conditions. In our experiments, although the $NO_3^-$ concentration in the PAW treatments was relatively high, we also detected elevated levels of nitrite ($NO_2^-$). Given that we observed significantly enhanced activities in other monitored physiological parameters following PAW treatments, we hypothesize that the elevated $NO_2^-$ concentration in our PAW may have mitigated the beneficial effects of nitrate, resulting in no significant changes in chlorophylls, carotenoids, or phenolic compounds in young bean leaves.

## 5. Conclusions

Plasma activated water (PAW) rich in RONS is an environmentally friendly technique that can replace chemical fertilizers and promote the growth of economically important crops such as common bean. In this unique study, we investigated the effects of different types of PAW-priming on common bean (*Phaseolus vulgaris*) under laboratory and real outdoor field conditions. We screened different types of PAW on germination in the laboratory. We used the most effective type of PAW (TS-1.5kHz 10m) for an outdoor experiment (lasting 64 days), where we tested the effect of PAW (priming, spraying, priming*spraying). The PAW-priming treatment proved to be the most effective treatment for plant length and their total biomass at the end of the experiment. The combination of priming*spraying showed an interesting trend in the physiological parameters of young leaves. Plants treated with PAW spraying showed higher SOD activity compared to the control. Based on our results, we hypothesize that PAW application could be beneficial in nitrogen-deficient soils and in environments with stress factors. Our further experiments will be directed towards a more detailed investigation of common bean growth in outdoor condition.

**Acknowledgements**

The research was supported by the EU NextGenerationEU through the Recovery and Resilience Plan for Slovakia under the projects No. 09I03-03-V02-00036 and No. 09I03-03-V03 00033 EnvAdwice.

**Conflicts of Interest**

The authors declare no conflicts of interest.

**Data Availability Statement**

The data that support the findings of this study are available from the corresponding author upon reasonable request.

by plasma activated tap water, demineralized water and liquid fertilizer," *RSC Adv.*, vol. 7, no. 50, pp. 31244–31251, Jun. 2017, doi: 10.1039/c7ra04663d.